%% file: main-sigconf.tex
\documentclass[sigconf]{acmart}
\usepackage{epsfig}
\usepackage{graphicx}
\usepackage{algorithm}
\usepackage{rotating}

\usepackage{tabularx}
\usepackage{subfigure}
\usepackage{booktabs}
\usepackage{bm}
\usepackage{multirow}

\AtBeginDocument{%
  \providecommand\BibTeX{{%
    \normalfont B\kern-0.5em{\scshape i\kern-0.25em b}\kern-0.8em\TeX}}}

\copyrightyear{2021}
\acmYear{2021}
\setcopyright{acmcopyright}\acmConference[SIGIR '21]{Proceedings of the 44th International ACM SIGIR Conference on Research and Development in Information Retrieval}{July 11--15, 2021}{Virtual Event, Canada}
\acmBooktitle{Proceedings of the 44th International ACM SIGIR Conference on Research and Development in Information Retrieval (SIGIR '21), July 11--15, 2021, Virtual Event, Canada}
\acmPrice{15.00}
\acmDOI{10.1145/3404835.3463010}
\acmISBN{978-1-4503-8037-9/21/07}

\acmSubmissionID{}

\begin{document}
\fancyhead{}

\title{Transfer-Meta Framework for Cross-domain Recommendation to Cold-Start Users}

\author{Yongchun Zhu$^{1,2,3}$, Kaikai Ge$^{3}$, Fuzhen Zhuang$^{4,5,*}$, Ruobing Xie$^{3}$, Dongbo Xi$^{1,2}$, Xu Zhang$^3$, Leyu Lin$^3$ and Qing He$^{1,2}$}
\affiliation{%
 \institution{$^1$Key Lab of Intelligent Information Processing of Chinese Academy of Sciences (CAS), Institute of Computing Technology, CAS, Beijing 100190, China}} 
\affiliation{%
\institution{$^2$University of Chinese Academy of Sciences, Beijing 100049, China}} 
\affiliation{%
\institution{$^3$WeChat Search Application Department, Tencent, China.}}
\affiliation{%
\institution{$^4$Institute of Artificial Intelligence, Beihang University, Beijing 100191, China.}} 
\affiliation{%
\institution{$^5$Xiamen Data Intelligence Academy of ICT, CAS, China.}} 
\affiliation{\{zhuyongchun18s, xidongbo17s, heqing\}@ict.ac.cn, \{kavinge, ruobingxie, xuonezhang, goshawklin\}@tencent.com,zhuangfuzhen@buaa.edu.cn}
\thanks{*Fuzhen Zhuang is the corresponding author.}

\renewcommand{\shortauthors}{Y. Zhu et al.}


\begin{abstract}
Cold-start problems are enormous challenges in practical recommender systems. One promising solution for this problem is cross-domain recommendation (CDR) which leverages rich information from an auxiliary (source) domain to improve the performance of recommender system in the target domain. In these CDR approaches, the family of Embedding and Mapping methods for CDR (EMCDR) is very effective, which explicitly learn a mapping function from source embeddings to target embeddings with overlapping users. However, these approaches suffer from one serious problem: 
the mapping function is only learned on limited overlapping users, and the function would be biased to the limited overlapping users, which leads to unsatisfying generalization ability and degrades the performance on cold-start users in the target domain.
With the advantage of meta learning which has good generalization ability to novel tasks, we propose a transfer-meta framework for CDR (TMCDR) which has a transfer stage and a meta stage. 
In the transfer (pre-training) stage, a source model and a target model are trained on source and target domains, respectively. In the meta stage, a task-oriented meta network is learned to implicitly transform the user embedding in the source domain to the target feature space. 
In addition, the TMCDR is a general framework that can be applied upon various base models, e.g., MF, BPR, CML. By utilizing data from Amazon and Douban, we conduct extensive experiments on 6 cross-domain tasks to demonstrate the superior performance and compatibility of TMCDR.
\end{abstract}




\maketitle

\input{Introduction}
\input{Model}

\input{Experiment}
\input{Conclusion}

\begin{acks}
The research work is supported by the National Key Research and Development Program of China under Grant No. 2018YFB1004300, the National Natural Science Foundation of China under Grant No. 61773361, U1836206, U1811461. 
\end{acks}

\bibliographystyle{ACM-Reference-Format}
\bibliography{main-sigconf}

\end{document}

%% file: Introduction.tex
\section{Introduction}
In the era of information explosion, how to efficiently obtain useful information from massive data is very important. Recommender systems play a major role in alleviating information overload. However, it is difficult to make cold-start recommendations, e.g., new users (user cold-start) and new items (item cold-start). Cross-domain recommendation (CDR)~\cite{pan2010transfer,man2017cross,kang2019semi} is a promising solution to address the cold-start problem.

CDR which leverages rich information from an auxiliary (source) domain to improve the performance of recommender system in the target domain has gained increasing attention in recent years. Actually, most CDR methods~\cite{singh2008relational,pan2010transfer,he2018general,gao2019cross,xi2020graph} aim to improve the overall performance for a target domain with the help of a source domain. Other methods address the cold-start problem which is more technically challenging as well as has great values from a practical perspective~\cite{mirbakhsh2015improving}. To address the problem, the Embedding and Mapping approach for CDR (EMCDR)~\cite{man2017cross} is very effective, which encodes user’s preferences of source and target domains on items into two embeddings, respectively, and then \textbf{explicitly} learns a mapping function from source embeddings to target embeddings with overlapping users. In other words, it minimizes the distance between the target embedding and the approximated embedding mapped from the source embedding for each overlapping user with Mean Squared Error (MSE) loss. With the advantage of EMCDR, many EMCDR-based approaches~\cite{zhu2018deep,fu2019deeply,kang2019semi} have been proposed. 

However, most of the EMCDR-based methods suffer from one serious problem. These methods explicitly learn the mapping function by minimizing the distance between the target embedding and the mapped embedding of the overlapping users. In other words, the number of training samples to learn the mapping function is equal to the number of overlapping users. In practice, the average ratio of the overlapping users to total users of any two domains is very low~\cite{kang2019semi}, e.g., in our experiments, the cross-domain tasks have 37388 overlapping users at most and 894 at least. Hence, the mapping function would be biased to the limited overlapping users, which leads to unsatisfying generalization ability and degrades the performance of the model on cold-start users in the target domain.

Meta learning~\cite{andrychowicz2016learning,finn2017model} has good generalization ability to novel tasks by training the model on lots of similar training tasks. With the advantage of meta learning, we propose a transfer-meta framework for CDR (TMCDR) which can replace the training procedure of most EMCDR-based methods~\cite{man2017cross,zhu2018deep,fu2019deeply,kang2019semi} and be applied upon various base models, e.g., MF~\cite{koren2009matrix}, BPR~\cite{rendle2009bpr}, CML~\cite{hsieh2017collaborative}. In detail, the proposed TMCDR has a transfer stage and a meta stage. 

\textbf{Transfer stage}: The transfer stage is similar to the embedding step in EMCDR~\cite{man2017cross}. However, the embedding step in EMCDR learns a source model and a target model with the user-item interaction of only overlapping users from scratch. However, with limited overlapping users, it is hard to learn the embedding of items which has not been interacted by the overlapping users.
In practice, each domain has a unique model for the overall recommendation. To address the problem, the unique model is directly utilized as pre-trained model. Thus, this stage is called as transfer stage. Note that the models should be embedding-based methods~\cite{koren2009matrix,rendle2009bpr,he2017neural}. 

\textbf{Meta stage}: The main idea of the meta stage is that a task-oriented meta network is learned to \textbf{implicitly} transform the overlapping user embedding in the source domain to the target feature space. The optimization goal of the cold-start task in CDR is learning knowledge from overlapping users and generalizing to cold-start users in the target domain. Inspired by meta learning (learning to learn), we construct special training tasks to simulate the target task. Each training task consists of a learning phase and a cold-start phase. The learning phase learns knowledge from one overlapping user. The cold-start phase uses another overlapping user to simulate a cold-start user. With the advantage of the popular Model-Agnostic Meta-Learning (MAML)~\cite{finn2017model}, we proposed a gradient-based training method to optimize a meta network. Different from the mapping function of EMCDR, the meta network is task-oriented, which denotes that the optimization goal is rating or ranking, not mapping. 

The main contributions of this work are summarized into three folds:
(1) To solve the cold-start problem in CDR, we propose a novel transfer-meta framework (TMCDR), which can be applied on most EMCDR-based methods with various base models, e.g., MF, BPR, CML.
(2) The proposed method is easy to implement in the online cold-start setting. With existing pre-trained models, once the task-oriented meta network is trained, it can be exploited for cold-start users.

%% file: Model.tex
\section{Model}

\subsection{Transfer Stage}
In practice, each domain has a unique model for the overall recommendation, and we use the source and target models as pre-trained models. The pre-trained model is trained on all data. Compared with the embedding model of EMCDR trained on samples of overlapping users, the pre-trained model has two advantages: 1) with more data, the model is more robust, which is hard to be bothered by data noise. 2) In EMCDR, the model can only learn embeddings of items interacted by the overlapping users. In contrast, with all data, the pre-trained model can capture information from all items. Hence, our TMCDR directly utilizes the pre-trained model. In our experiments, we simulate the pre-training procedure, i.e., a model is trained with all data as the pre-trained model. In practice, various models are exploited for different applications. To testify our TMCDR is compatible with various models, we implement four popular embedding models into TMCDR framework, including MF~\cite{koren2009matrix}, BPR~\cite{rendle2009bpr}, ListRank-MF~\cite{shi2010list}, and CML~\cite{hsieh2017collaborative}.
Note that we have defined a rating matrix $R \in \{0, 1\}$ above, and the problem is a binary classification task.

\begin{algorithm} [t] 
    \caption{Transfer-Meta framework for CDR (TMCDR)}\label{alg}
    \flushleft{\textbf{Input}: Given user and item sets of source and target domains, $U^s, U^t, V^s, V^t$. The overlapping user set $U^o$. The rating matrix $R^s, R^t$. 
    
    \textbf{Input}: Task-oriented meta network $f_\theta$.
    
    \textbf{Input}: The step size (learning rate) $\lambda, \alpha$.
    
    \textbf{Transfer Stage}:
    
    \begin{itemize}
        \item A pre-trained source model contains $\mathbf{u}^s, \mathbf{v}^s$.
        \item A pre-trained target model contains $\mathbf{u}^t, \mathbf{v}^t$.
    \end{itemize}

    \textbf{Meta Stage}: utilize the source embedding of overlapping users $\mathbf{u}^s$ and the target item embedding $\mathbf{v}^t$ to optimize the task-oriented meta network $f_\theta$.
    
    \begin{enumerate}
        \item randomly initialize $\theta$.
        \item \textbf{while} not converge \textbf{do}:
        \item \quad sample batch of user groups \{$U_1, ..., U_n$\} from $U^o$.
        \item \quad \textbf{for} $U_i \in \{U_1, ..., U_n\}$ do:
        \item \quad \quad divide $U_i$ into two disjoint sets $U_a, U_b$
        \item \quad \quad define two training sets $D_a, D_b$ with $U_a, U_b$
        \item \quad \quad evaluate loss $\mathcal{L}_\theta$ with $D_a$
        \item \quad \quad compute updated parameter $\theta' = \theta - \lambda \frac{\partial \mathcal{L}_\theta}{\partial \theta}$
        \item \quad \quad evaluate loss $\mathcal{L}_{\theta'_i}$ with $D_b$
        \item \quad update $\theta = \theta - \alpha \sum_{U_i \in \{U_1, ..., U_n\}} \frac{\partial \mathcal{L}_{\theta'_i}}{\partial \theta}$
        \item \textbf{end while}
    \end{enumerate}
    \textbf{Test Stage}: for a cold-start user $u$, we use $f_\theta(\mathbf{u}^s)$ as the user embedding for prediction.}
\end{algorithm}

\subsection{Meta Stage}
After the transfer stage, we can obtain the pre-trained source and target models (the users' and items' embeddings $\mathbf{u}^s, \mathbf{v}^s, \mathbf{u}^t, \mathbf{v}^t$). With the pre-trained embeddings fixed, we put forward a task-oriented meta network that can implicitly transform the source embeddings into the target feature space.   

Recall that the main idea of cold-start in CDR is to learn knowledge from overlapping users and generalizing to cold-start users in the target domain. Meta learning~\cite{andrychowicz2016learning,finn2017model} has good generalization ability to novel tasks by training the model on a variety of similar training tasks. Inspired by meta learning, to simulate the target task, we construct similar training tasks, and each training task consists of a learning phase and a cold-start phase.

The goal of the learning phase is to simulate that learning knowledge from overlapping users. Hence, each learning phase contains all user-item interaction samples of an overlapping user. In addition, the cold-start phase is to simulate the cold-start users in the target domain. However, the real cold-start users have no interaction behavior in the target domain, and the model cannot directly learn knowledge from these users. Thus, we utilize another overlapping user to simulate a cold-start user. 

Each training task only has training samples of two users. This suffers from a challenge that a user could have very limited interaction samples. We find that it may lead to unstable training. In ~\cite{pan2019warm}, they use users whose number of interaction samples exceeds a certain threshold for training. In this paper, we propose a group strategy that both the learning phase and cold-start phase contain several users, but the users of the two phases are disjoint. We denote the users of the learning phase and cold-start phase as $U_a$ and $U_b$, respectively. Two training sets in the target domain with all samples of $U_a$ and $U_b$ are denoted as $D_a$ and $D_b$, respectively. 

\begin{table}[t]
  \centering
  \caption{Recommendation performance on 6 CDR tasks. $*$ indicates $0.05$ level, paired t-test of TMCDR\_MF vs. the best baselines.}
    \begin{tabular}{c||cc||cc}
    \toprule
    \multirow{2}[2]{*}{Method} & AUC   & NDCG@10 & AUC   & NDCG@10 \\
          & \multicolumn{2}{c||}{Scenario1} & \multicolumn{2}{c}{Scenario2} \\
    \midrule
    CMF   & 0.6490 & 0.1696 & 0.6996 & 0.2076 \\
    BPR   & 0.7226 & 0.2182 & 0.7160 & 0.2379 \\
    ListRank-MF & 0.6648 & 0.1709 & 0.7232 & 0.2204 \\
    CML   & 0.6470 & 0.1408 & 0.6986 & 0.2147 \\
    CST   & 0.7240 & 0.2137 & 0.7124 & 0.2324 \\
    SSCDR & 0.7245 & 0.0089 & 0.6745 & 0.0013 \\
    EMCDR\_MFori & 0.6942 & 0.1978 & 0.6511 & 0.1747 \\
    EMCDR\_MF & 0.7271 & 0.2103 & 0.6923 & 0.1985 \\
    TMCDR\_MF & \textbf{0.7501*} & \textbf{0.2246*} & \textbf{0.7253*} & \textbf{0.2427*} \\
    \midrule
          & \multicolumn{2}{c||}{Scenario3} & \multicolumn{2}{c}{Scenario4} \\
    \midrule
    CMF   & 0.7769 & 0.3066 & 0.7295 & 0.2349 \\
    BPR   & 0.7737 & 0.3065 & 0.7199 & 0.2150 \\
    ListRank-MF & 0.7640 & 0.2902 & 0.7409 & 0.2277 \\
    CML   & 0.8191 & \textbf{0.3548} & 0.7857 & 0.2647 \\
    CST   & 0.7995 & 0.2960 & 0.7842 & 0.2563 \\
    SSCDR & 0.7956 & 0.3080 & 0.6545 & 0.1628 \\
    EMCDR\_MFori & 0.7273 & 0.2284 & 0.7307 & 0.1990 \\
    EMCDR\_MF & 0.8011 & 0.3055 & 0.7936 & 0.2670 \\
    TMCDR\_MF & \textbf{0.8282*} & 0.3334 & \textbf{0.8056*} & \textbf{0.2775*} \\
    \midrule
          & \multicolumn{2}{c||}{Scenario5} & \multicolumn{2}{c}{Scenario6} \\
    \midrule
    CMF   & 0.8465 & 0.3420 & 0.8339 & 0.3764 \\
    BPR   & 0.8108 & 0.3283 & 0.8138 & 0.3659 \\
    ListRank-MF & 0.8136 & 0.3106 & 0.8191 & 0.3281 \\
    CML   & 0.8466 & 0.3409 & 0.8405 & 0.3707 \\
    CST   & 0.8524 & 0.3405 & 0.8406 & 0.3742 \\
    SSCDR & 0.8144 & 0.2925 & 0.8317 & 0.3644 \\
    EMCDR\_MFori & 0.7307 & 0.1990 & 0.7627 & 0.2703 \\
    EMCDR\_MF & 0.8438 & 0.3322 & 0.8297 & 0.3702 \\
    TMCDR\_MF & \textbf{0.8589*} & \textbf{0.3483*} & \textbf{0.8442*} & \textbf{0.3778*} \\
    \bottomrule
    \end{tabular}%
  \label{tab:result}
\end{table}%

We define meta network as $f_\theta(\cdot)$, and $\theta$ denotes the parameters. Besides, $f_\theta(\mathbf{u}^s_i)$ represents the transformed embedding of $\mathbf{u}^s_i$. The meta network should be optimized on lots of training tasks. We define the loss function to be the same as the optimization goal of the pre-training task, so we call it task-oriented loss. The pre-training task could be one of MF~\cite{koren2009matrix}, BPR~\cite{rendle2009bpr}, ListRank-MF~\cite{shi2010list}, and CML~\cite{hsieh2017collaborative} as mentioned above. \textbf{Task-oriented loss utilizes the transformed user embedding $f_\theta(\mathbf{u}^s_i)$ not $\mathbf{u}^t_i$}. The overall training procedure of the task-oriented meta network is following the meta learning paradigm~\cite{finn2017model}. Firstly, the loss of the learning phase can be formulated as:
\begin{equation}
\begin{split}
    \mathcal{L}_\theta = \sum_{x \in D_a} L_\textit{task}(x),
\end{split}
\end{equation}
where $\textit{task} \in \textit{MF}, \textit{BPR}, \textit{ListRank-MF}, \textit{CML}$, and $x \in D_a$ denotes one sample. By computing the gradient of $\mathcal{L}_\theta$ and taking a step of gradient descent, we get a new adapted parameter:
\begin{equation}
    \theta' = \theta - \lambda \frac{\partial \mathcal{L}_\theta}{\partial \theta},\label{eq:10}
\end{equation}
where $\lambda > 0$ is the step size of gradient descent (learning rate). Now that we have a new parameter $\theta'$ which is trained with the overlapping users $U_a$, and we can test the adapted model $f_{\theta'}$ on the cold-start users $U_b$. Similarly, the loss of the cold-start phase is:
\begin{equation}
\begin{split}
    \mathcal{L}_{\theta'} = \sum_{x \in D_b} L_\textit{task}(x).
\end{split}
\end{equation}
Then we minimize the $\mathcal{L}_{\theta'}$ to update $\theta$:
\begin{equation}
\begin{split}
    \theta &= \theta - \alpha \frac{\partial \mathcal{L}_{\theta'}}{\partial \theta}\\
\end{split}
\end{equation}
where $\frac{\partial \theta'}{\partial \theta}$ can be computed by the Equation~(\ref{eq:10}). Note that the meta-optimization is performed over the model parameters $\theta$, whereas the objective is computed using the updated model parameters $\theta'$. Actually, the meta stage aims to optimize the parameters of task-oriented meta network such that one or a small number of gradient steps on a group of simulated cold-start users will produce maximally effective behavior on that the real-world cold-start users. 

Finally, we come to the overall training algorithm of TMCDR, which can update the meta-parameters by stochastic gradient descent in a mini-batch manner, see Algorithm~\ref{alg}.

%% file: Experiment.tex
\section{Experiments}
\subsection{Experimental Settings}
\textbf{Dataset}. Two real-world datasets are adopted for evaluation, Amazon\footnote{http://jmcauley.ucsd.edu/data/amazon/} and Douban\footnote{https://www.douban.com}. Both datasets have been used for the CDR problem.

The first dataset is a public Amazon dataset, which has various versions. And we use the Amazon-5cores dataset that each user or item has at least five ratings. The dataset contains 24 different item domains. Among them, we choose the seven popular categories: apps\_for\_android, video\_games, home\_and\_kitchen, movies\_and\_tv, cds\_and\_vinyl, books and tools\_and\_home\_improve-
ment. Then, we define four CDR scenarios as Scenario 1: apps\_for\_
android $\rightarrow$ video\_games, Scenario 2: home\_and\_kitchen $\rightarrow$ tools\_and
\_home\_improvement, Scenario 3: movies\_and\_tv $\rightarrow$ cds\_and\_vinyl, and Scenario 4: books $\rightarrow$ movies\_and\_tv.

The second dataset is Douban dataset, which includes three recommendation tasks like movies, music, and book recommendations. We utilize the Douban dataset to construct another two cross-domain tasks: Scenario 5: movie $\rightarrow$ music, and Scenario 6: music $\rightarrow$ book.

\textbf{Evaluation Protocol}: To evaluate the performance of the proposed framework on the CDR tasks for cold-start users, for each task, we randomly select about 20\% overlapping users as cold-start users, and use all samples of these users in target domain for the test stage. we adopt three standard metrics, AUC and NDCG@K, which are widely used in recommendation~\cite{man2017cross,gao2019cross} to evaluate the ranking performance of each method. 



\textbf{Baselines}: The baselines can be divided into two groups: single-domain and cross-domain. In the first group, we consider both source and target domains as a single domain and utilize popular CF method, including CMF~\cite{singh2008relational}, BPR~\cite{rendle2009bpr}, ListRank-MF~\cite{shi2010list}, and CML~\cite{hsieh2017collaborative}. The second group includes state-of-the-art CDR methods for cold-start users, including CST~\cite{pan2010transfer}, EMCDR~\cite{man2017cross}, and SSCDR~\cite{kang2019semi}.
Both EMCDR and our TMCDR are general frameworks for many embedding models. Thus, we apply EMCDR and TMCDR on MF, BPR, ListRank-MF, and CML, e.g., TMCDR\_MF.

\textbf{Implementation Details}: For simplicity, we intentionally transform the rating data into binary (1/0 indicate whether a user has interacted with an item or not) to fit the problem setting of implicit feedback following~\cite{gao2019cross}. In the training stage, for each positive sample, we randomly sample 4 negative samples. For all methods, we set the dimension of embedding as 256, and mini-batch size of 1280. We employ the Adam~\cite{kingma2014adam} optimizer and search its learning rate within $\{0.001, 0.002, 0.005, 0.01\}$. We fix $\lambda = 0.005$ in meta stage. For all EMCDR-based methods and TMCDR, we use a single fully connected layer as mapping function and meta network. We report the average AUC and NDCG@10 on three random trials.

\vspace{-0.2cm}
\subsection{Results}
\textbf{Recommendation Performance.} We demonstrate the effectiveness of TMCDR on six CDR tasks. The experimental results evaluated by AUC and NDCG@10 are shown in Table~\ref{tab:result}. The experimental results reveal several insightful observations. (1) With various CDR scenarios, TMCDR outperforms most compared methods which demonstrates the effectiveness of TMCDR. The improvement mainly comes from the task-oriented meta network.
(2) On all CDR scenarios, EMCDR\_MF largely outperforms EMCDR$_{ori}$\_MF. The main reason is that the overlapping users only cover a part of items, and the models cannot learn other items which have not been interacted by overlapping users. It also demonstrates that using all samples is more effective than only using samples of overlapping users. 
(3) We can find that CMF, BPR, ListRank-MF, and CML have different performances on various scenarios, which testifies different tasks should adopt different models. Especially in Scenario 3, CML outperforms all methods on NDCG@10.
(4) The confidence intervals of results on the Douban dataset are smaller than the Amazon dataset. We think the reason is that Douban dataset has more overlapping users, which makes the training process more stable.
(5) CST only utilizes the source model as a pre-trained model and regularizes the parameters can obtain remarkable results, which demonstrates fine-tuning recommendation models on a pre-trained model is effective.

\begin{figure}[t]
\centering
\begin{minipage}[b]{1\linewidth}
\centering
\includegraphics[width=0.95\linewidth]{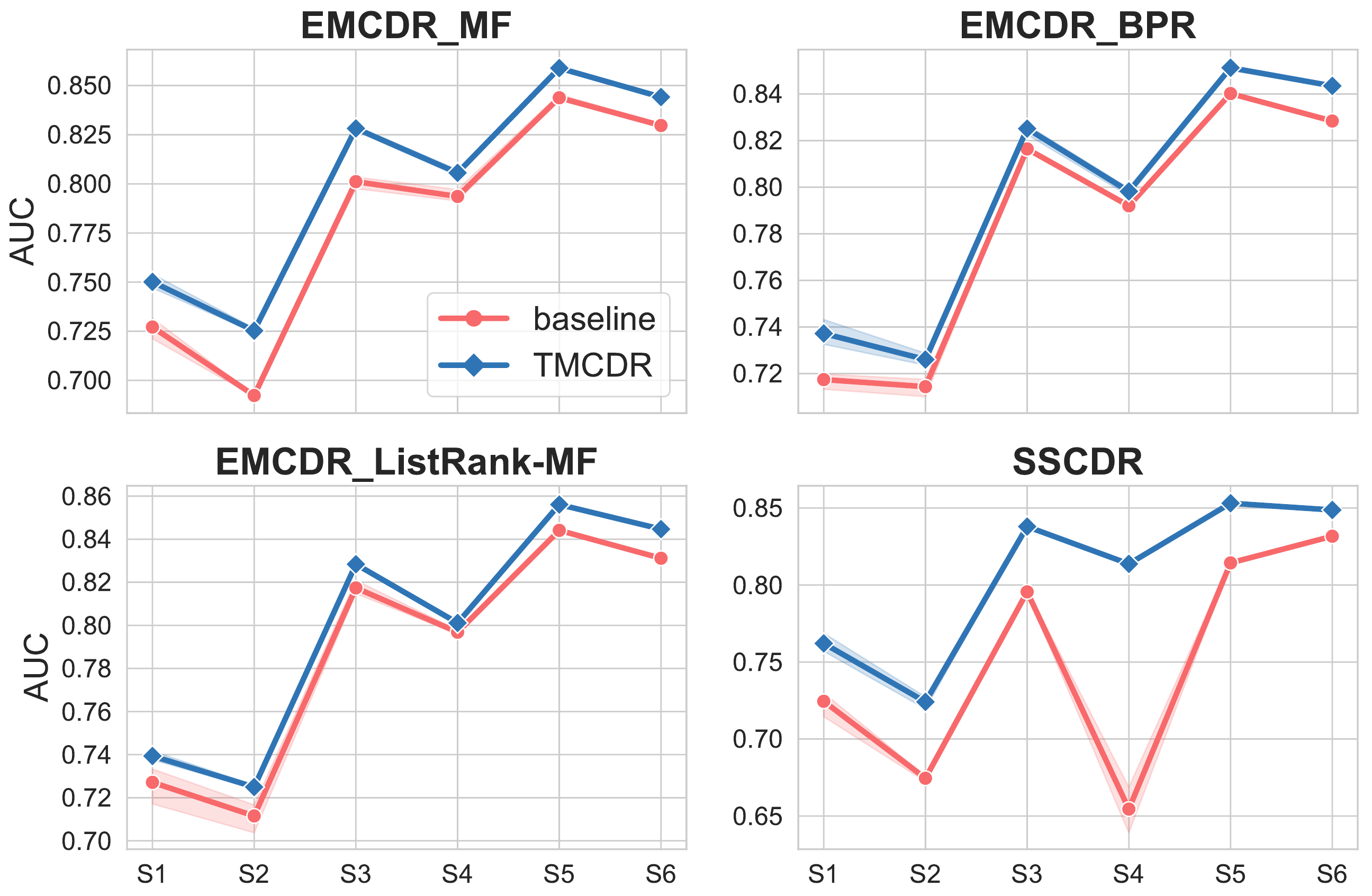}
\end{minipage}
\caption{Generalization experiments.}\label{fig}\vspace{-0.5cm}
\end{figure} 

\textbf{Generalization.} Our TMCDR is a general framework that can be applied on most EMCDR-based methods with various base models, e.g., MF, BPR, ListRank-MF, and CML. We compare four EMCDR-based methods, EMCDR\_MF, EMCDR\_BPR, EMCDR\_
ListRank-MF, and SSCDR. Note that SSCDR is a combination of CML and EMCDR, so we do not compare TMCDR with EMCDR\_CML. To implement our TMCDR, the mapping function of these methods is replaced by the task-oriented meta network.  The results are drawn in Figure~\ref{fig}, and the red lines denote the baselines (EMCDR-based methods), while the blue lines represent the modified methods (TMCDR). And, the bands are confidence intervals over three runs. From the Figure~\ref{fig}, we can find that on most tasks our TMCDR can improve the performance of various EMCDR-based methods, which demonstrates the generalization ability of the TMCDR.



%% file: Conclusion.tex
\vspace{-0.1cm}\section{Conclusion}
In this paper, we studied CDR to cold-start users from the perspective of meta learning. EMCDR is a popular approach in this area, and there are many EMCDR-based methods. However, with limited overlapping users, the mapping function would be overfitting. To address the problem, with the advantage of MAML, we proposed a novel Transfer-Meta Framework for CDR (TMCDR), which learns a task-oriented meta network.
Besides, TMCDR is a general framework that can be applied on most EMCDR-based methods with various base models, e.g., MF, BPR, CML. Finally, we conducted extensive experiments on real-world datasets collected from Amazon and Douban to validate the effectiveness and compatibility of our proposed TMCDR.